\documentclass{elsart}

\usepackage{amssymb}

\begin{document}

\begin{frontmatter}


% use the thanksref command within \title, \author or \address for footnotes:
\title{The Diverse Population of ULIRGs\thanksref{label1}}
\thanks[label1]{Support for this work was provided by NASA through
grant number GO-06346.01-95A from the Space
Telescope Science Institute, which is operated by
AURA, Inc., under NASA contract NAS5-26555.}
\author{Kirk D. Borne}
\address{Raytheon ITSS and NASA Goddard Space Flight Center, Greenbelt, MD
\thanksref{email}}
\thanks[email]{E-mail: Kirk.Borne@gsfc.nasa.gov}

\author{S. Arribas, H. Bushouse, L. Colina, \& R. A. Lucas}

\begin{abstract}
We present results from an on-going Hubble Space Telescope (HST) survey
of a large sample of ULIRGs (Ultra-Luminous IR Galaxies).  New
ground-based observations are now being used to complement the HST data
and to assist in the interpretation of these complex objects.  A rich
spectroscopic, morphological, and dynamical diversity is found within
the ULIRG population, nearly 100\% of which are merger and/or collision
remnants.  The consequences of this diversity may apply to the
interpretation of distant submm/FIR sources and their subsequent evolution.
\end{abstract}


\end{frontmatter}

% main text
\section{Collisions, Mergers, Starbursts, and ULIRGs}

It has been known since the pioneering work of Larson \& Tinsley \cite{lar}
that collisionally disturbed galaxies (e.g.,\cite{arp}) 
have abnormally high star formation rates compared to
isolated galaxies of similar type.  It was not until IRAS discovered
the population of luminous IR galaxies (LIRGs) and ultraluminous IR
galaxies (ULIRGs), having extremely high star formation rates
(100--1000$\times$ the Galactic star formation rate), that the links
between strong star formation, high IR luminosity, and galaxy-galaxy
encounters were made inseparable \cite{jos}.  The
observation that some galaxies have higher rates of star formation than
can be sustained by their current gas content over a full Hubble time
led to the idea of "starbursting" galaxies \cite{wee}.  
It has thus been unequivocally established that the
LIRG, ULIRG, starburst, and collision+merger processes are physically
related phenomena that are intimately connected to the star formation
history of galaxies, galaxy formation, and galaxy evolution (for a
review, see \cite{san}).  In fact, most recently, it has been estimated
that a significant portion of the cosmic IR background 
is produced by cosmologically distant LIRGs and ULIRGs 
(i.e., dusty starbursts; \cite{bar}, \cite{bla}, \cite{sma}).

\section{The Rich Diversity within the ULIRG Population}

We have been studying a large sample of ULIRGs with both HST imaging
and ground-based spectroscopy (\cite{arr}, \cite{bo97a}, \cite{bo97b},
\cite{bo98}, \cite{bo99a}, \cite{bo99b}, \cite{bo2000}, \cite{co99},
\cite{co2000}).  Among the plethora of results being derived from this
rich survey database, we have found strong evidence for a
multiple-merger origin for many of the ULIRGs in our sample
\cite{bo2000}.  We have established a morphological classification
scheme for ULIRGs that indicates that the sample is nearly equally
divided between single objects (e.g., merged; disturbed; or IR-luminous
QSOs) and multiple objects (e.g., pairs; compact groupings; or strongly
interacting multiples).  We find very little luminosity variation
across these morphological classes \cite{bo99a}.  We have also verified
the long-known belief that the ULIRG population has a high interaction
rate.  From a sample of nearly 130 ULIRGs, we find $\sim$98\% show
evidence for close neighbors, tidal disturbances, or on-going merging.
In several cases, the galaxies were previously classified as isolated
and/or undisturbed from low-resolution ground-based imaging.  Our
sample is the largest that has been used to derive this interaction
rate estimate.  Most recently, we have been obtaining multi-fiber
integral spectroscopy for several ULIRGs, which indicate that luminous
gas-rich knots are ubiquitous among these galaxies.  In every case, we
find multiple line emission sources, frequently in regions detached
from the cores of the galaxies or from any other region that is
luminous in continuum light.  These line-emitting regions have spectral
characteristics of H\thinspace II regions, LINERS, or AGN.  In some
cases, these line-emitting clouds may be simply reflecting the emission
from a dust-obscured nuclear source.  In the case of Mrk~273, we find a
LINER nucleus and an extended off-nucleus Seyfert 2 nebula
\cite{co99}.  In the case of IRAS~12112$+$0305, we find that the
observed ionized gas distribution is decoupled from the stellar main
body of the galaxy, with the dominant continuum and emission-line
regions separated by projected distances of up to 7.5 kpc.  The two
optical nuclei are detected as faint emission-line regions, and their
optical properties are consistent with being dust-enshrouded weak
[O\thinspace I] LINERs \cite{co2000}.  In the case of
IRAS~08572$+$3915, we find no evidence for a LINER or Seyfert-like
nucleus in either of the galaxies, contrary to previous claims.  This
is unusual for a {\it{warm}} ULIRG such as 08572$+$3915.
Tidal-induced star-forming knots, $\sim$7~kpc from the nuclei and along
the tidal tails, are traced by the presence of bright [O\thinspace III]
emission \cite{arr}.

In summary, we find that the ULIRG population of galaxies has a rich
dynamical diversity, demonstrated both morphologically and
spectroscopically.  This points to a rich evolutionary history for
these objects, with strong connections between their hierarchical mass
assembly history, dust and gas evolution, star formation episodes, 
metal enrichment, and nuclear activity.  The consequences of these
connections are found in the spectral energy distributions both of these
galaxies and of their high-redshift counterparts, and thus our
interpretations of these connections for nearby galaxies may thus also
be applicable to the interpretation of distant submm/FIR sources now
being discovered and soon to be discovered in the new long-wavelength
ultra-sensitive galaxy surveys.

% Bibliographic references with the natbib package:
% Parenthetical: \citep{Bai92} produces (Bailyn 1992).
% Textual: \citet{Bai95} produces Bailyn et al. (1995).
% An affix and part of a reference:
%   \citep[e.g.][Ch. 2]{Bar76}
%   produces (e.g. Barnes et al. 1976, Ch. 2).

\end{document}